# Automated Estimation of Equivalent Circuit Model from Impedances with Long Short-Term Memory

Ryoma Iki [a, *], Motoya Furugori [a], Noboru Katayama [b]
[a]Department of Electrical Engineering, Graduate School of Science and Technology, Tokyo University of Science, 2641 Yamazaki, Noda, Chiba, 278-8510 Japan
[b]Department of Electrical Engineering, Faculty of Science and Technology, Tokyo University of Science, 2641 Yamazaki, Noda, Chiba, 278-8510 Japan
* 7325509@ed.tus.ac.jp

**Abstract**
Electrochemical Impedance Spectroscopy (EIS) is a widely used, non-destructive technique for characterizing electrochemical systems, and its analysis typically relies on fitting the measured spectra to an Equivalent Circuit Model (ECM). Selecting an appropriate ECM, however, remains a major bottleneck: knowledge-based selection requires expert judgment and is difficult to reproduce, while existing automated approaches either choose from a fixed set of candidate circuits or, in the case of Gene Expression Programming, require repeated equivalent-circuit fitting and a predetermined circuit scale. Here, we propose a machine learning method that estimates an ECM directly from an impedance spectrum by representing the circuit as a serialized string of symbols and generating this string with a Long Short-Term Memory (LSTM) network coupled to a convolutional feature extractor. Because the LSTM inherently handles variable-length sequences, the method produces the circuit topology directly, without any fitting during estimation nor prior assumption for the number of elements. A fourth-root transformation of the impedance is introduced to emphasize the mid-frequency features essential for distinguishing circuits, and an adaptive beam search yields multiple ranked candidates. Evaluated on 100,000 synthetic datasets generated from 119 circuit topologies with 1% added noise on impedances, the method identified the correct topology as the most probable ECM in 77.8% of cases and among the top five candidates in 98.8% of cases, with an average estimation time of 17.8 milliseconds per dataset—several orders of magnitude faster than reported fitting-based approaches. These results indicate that direct topology generation with a neural network is a promising route toward fully automated, expert-independent ECM estimation.

# 1. Introduction

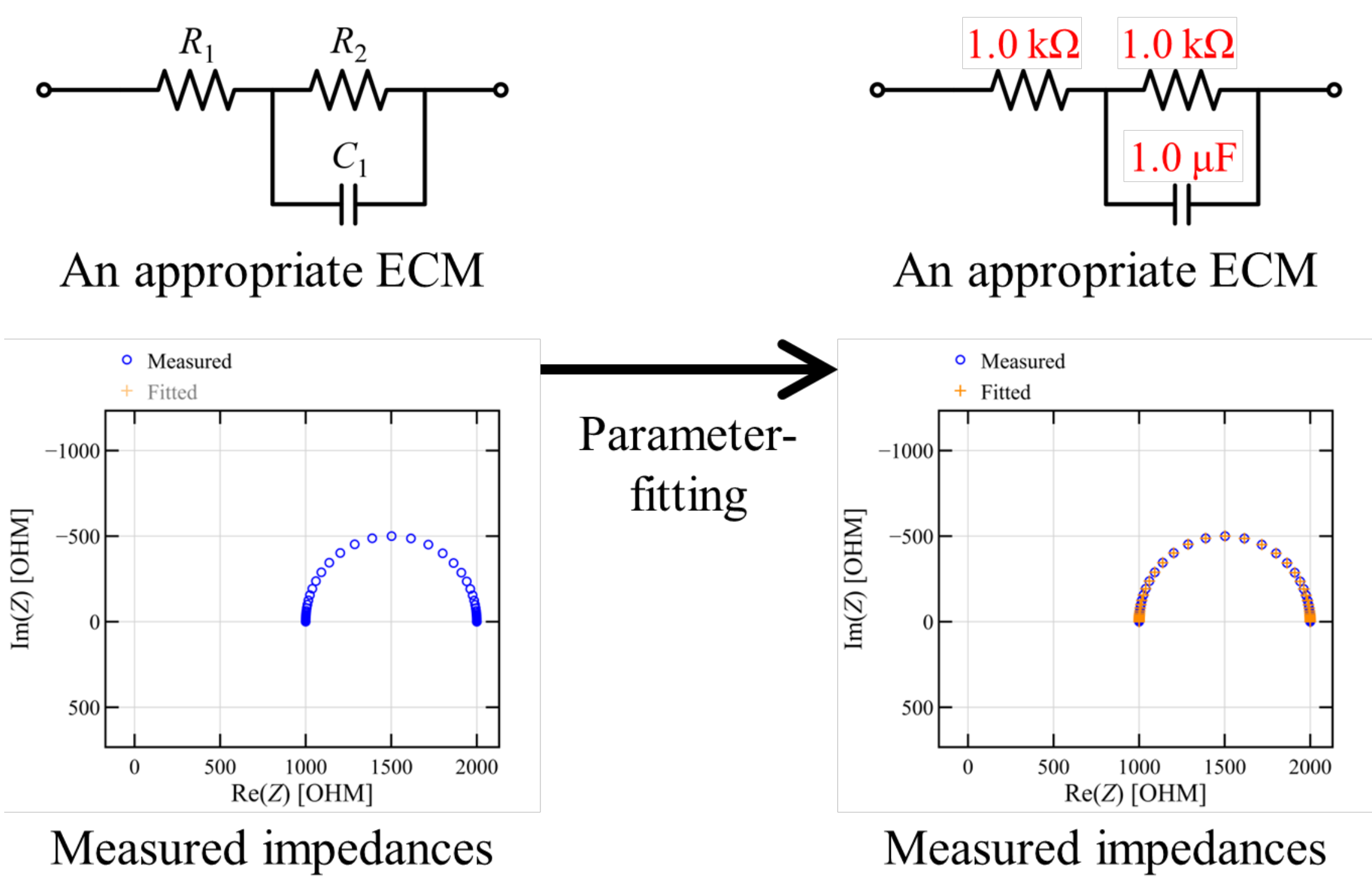

Fig. 1. Example of parameter fitting

Electrochemical Impedance Spectroscopy (EIS) is one of the most frequently applied analytical tools for a variety of electrochemical devices such as batteries [1][2][3] and fuel cells [4] and is often applied for an analysis of electrocatalysts and corrosion [5]. In EIS measurement, a microscopic AC signal with a swept frequency is applied to an object and at each frequency the impedance can be calculated from the amplitude ratio and phase difference between the voltage and current. The remarkable feature of EIS is that it is non-destructive, that is, the measurement can be done without damaging the object or changing its characteristics.

The goal of EIS is often to determine an Equivalent Circuit Model (ECM) appropriate for the object, especially in the field of electrochemistry. The measured impedances and an appropriate ECM provide the parameters of the ECM through a process called parameter fitting as shown in Fig. 1. The parameters of ECM are determined by performing optimization to minimize the difference between the measured and the fitted impedances. Parameter fitting is widely applied especially in the field of electrochemistry for analyzing electrochemical phenomena such as ion diffusion and charge transfer [6].

To perform parameter fitting, an appropriate ECM must first be estimated, and several approaches exist for deciding which ECM to be applied. The first is a knowledge-and-experience-based approach, in which an ECM is estimated by electrochemistry experts based on their knowledge, experiences, and previous studies. This approach has several drawbacks: the estimation is too difficult for non-experts, estimated ECMs sometimes differ depending on experts, and there are many cases in which even experts have difficulty owing to the complexity of the impedances. Therefore, it is expected to develop an automated way independent to experts.

Another is an automated methodology based on Distribution of Relaxation Times (DRT), which estimates how many resistive-capacitive elements—each a parallel combination of a resistor and a capacitor (or sometimes a constant phase element)—should be connected in series from the number of peaks seen in the graph of impedance spectrum. In advanced technology called generalized DRT [7], not only resistive-capacitive but also resistive-inductive elements and lumped elements such as resistive, inductive, and capacitive elements can also be estimated. This methodology relies on little decisions by experts and rarely differs among experts, however, it still has the issue that ECMs with only resistive-capacitive, resistive-inductive, or lamped elements can be estimated and no ECMs including any other elements can be estimated.

To deal with this issue, another automated way based on parameter fitting was proposed in our previous study [9]. In this methodology, constraints of estimation, such as the maximum number of elements in ECMs and the kind of elements available in ECMs, are designated before the estimation. Then, all circuit models that satisfy the designated constraints are generated, and parameter fitting is performed for each circuit model. Theoretically, an appropriate designation of the constraints and ideal non-limited calculation resources such as calculation time and machine memories would enable us to gain all possible ECMs. However, under realistic limited calculation resources, enormous time was required for the estimation of complex ECMs.

A further approach is Machine Learning (ML)-based automated methodology, which has the potential to automatically estimate an appropriate ECM from a variety of ECM candidates with realistic calculation resources. Given a large number of datasets, ML model automatically extracts hidden features in the datasets and relationship between input and output, bypassing the need for explicit algorithmic instructions. Some previous researches have also applied ML-based methodology: one is to apply an ML model such as a classification[8][10][11] model or genetic algorithm[1] to find an appropriate ECM from some limited number of candidates. This ML-based methodology specializes for some limited applications: for example, if the candidates are collected from ECMs commonly seen in lithium-ion batteries, this methodology specializes in the application for lithium-ion batteries and usually inappropriate for any other devices since any ECMs not listed in the candidates cannot be estimated. This limitation is mainly because of structure of an ML model: a classification model and genetic algorithm restrict outputs to a finite set of discrete categories and their combinations, respectively, making it difficult to estimate complex targets with a large variation of structural combinations or unknown topologies.

Another ML-based methodology applies Gene Expression Programming (GEP) [12][13][14], which has the potential to be applied to a wider range of objects. GEP is an evolutionary algorithm for the automated discovery of complex functional structures that integrates the structural flexibility of genetic programming with the efficient linear genetic encoding of gene expression. This GEP-based methodology selects, as the first generation, several strings (each a sequence of characters, called a chromosome in GEP) representing ECMs, either randomly generated or selected from strings representing well-known ECMs for various applications. The next generation of strings is created by applying genetic operators to each string in the current generation—for example, replacing one character with another (mutation), overwriting part of a string with part of another (recombination), or moving a substring from one position to another (transposition). Parameter fitting is performed for each ECM represented by strings in a generation, and those with better fitting results are used for creating next generation. This GEP-based methodology is truly novel in that it has the potential to generate any kind of ECMs under the given constraints such as kind of elements or the maximum number of elements that can be contained in an ECM. It has been reported that a modified Randles circuit model could be found after a few trials of this methodology [12]. However, a major issue remains with GEP-based methodology: the ECM estimation is indirect. Parameter fitting is required after each generation in order to determine validity of each candidate as ECM, which causes substantial consumption of calculation resources. Also, the scale of an ECM must be assumed prior to a GEP-based prediction since the maximum number of elements that an ECM can contain is limited by the architecture of GEP model.

To achieve a direct estimation technique with a variable ECM scale, it is worth considering the introduction of a Neural Network (NN) model, another type of ML model. An NN model automatically captures features or rules hidden in a given dataset by tuning its internal parameters such as gradients or bias. Once an NN model is well trained with on many datasets, it can estimate hidden features or rules directly from given data. As an advanced types of neural networks capable of outputting sequential data, such as string or a list of numbers that follows a specific rule, a Long Short-Term Memory (LSTM) network has recently attracted significant attention. LSTM is a specialized type of Recurrent Neural Network (RNN), which is neural network with a feedback loop to model sequential data. Compared with a typical RNN, LSTM can process complex sequential data more efficiently owing to its structure, which contains not only a gated architecture as hidden state to selectively memory short-term contexts but also a cell state to selectively retain long-term contexts. Owing to its ability to output sequential data, LSTM is widely applied in a variety of fields, such as image captioning [15][16], polymetric science [17], fake news detection [18], and language translation [19].

In this study, we propose using LSTM for ECM estimation. Since LSTM can output variable-length sequential data, such as natural language, as the output of a neural network, we attempt to exploit this capability to output ECMs. However, an ECM is composed of a topological structure in which circuit elements are connected in series and parallel; we therefore represent this topological structure as sequential data. Specifically, an ECM is expressed using symbols for the circuit elements and for series and parallel connections, and is serialized into an expression-like form. This allows a wide variety of ECMs to be represented in a form that LSTM can handle. Moreover, the representation is easily extensible by increasing the number of symbol types. In addition, another fundamental advantage of the proposed method is that, after the input impedance is interpreted by a convolutional neural network as a feature extractor, the subsequent LSTM can directly output an ECM. Unlike GEP-based methods[20][21], which require equivalent-circuit fitting at every feedback iteration to evaluate the ECM being generated, the proposed method does not require such a process. This leads to a substantial reduction in the computation time required for an estimation.

The remainder of this paper is organized as follows. Section 2 describes the proposed ML model and how this model is applied and evaluated. Section 3 describes the methodology and conditions of the training and evaluation processes. Section 4 presents the results of the evaluation, and section 5 concludes this paper.

# 2. Proposed methodology

This section describes the proposed methodology. ML generally consists of two processes: the training process, which trains an ML model, and the evaluation process, which assesses whether the model is valid for unseen input data or applies the model to unknown data to an estimation result. Trained using on a large number of datasets consisting of ECMs and their impedances at swept frequencies, a NN model such as LSTM has the potential to estimate an appropriate ECM even from an unseen impedance. In this research, we employed an LSTM to directly generate outputs of variable length with no prior assumptions about the scale of an ECM. Section 2.1 describes the structure of the proposed ML model, which combines a feature extractor and LSTM, as well as a preprocess method that improves the likelihood of generating an appropriate ECM by making it easier for the model to analyze features the feature contained in mid-frequency range. As another strategy for improving likelihood of generating an appropriate ECM, an adaptive beam search strategy that generates multiple ECM candidates is also employed, as described in Section 2.2. As described in section 2.3, each of the large datasets, ECM is generated by randomly generating an ECM under a specific rule and calculating its impedances at swept frequencies.

## 2.1 Structure of machine learning model

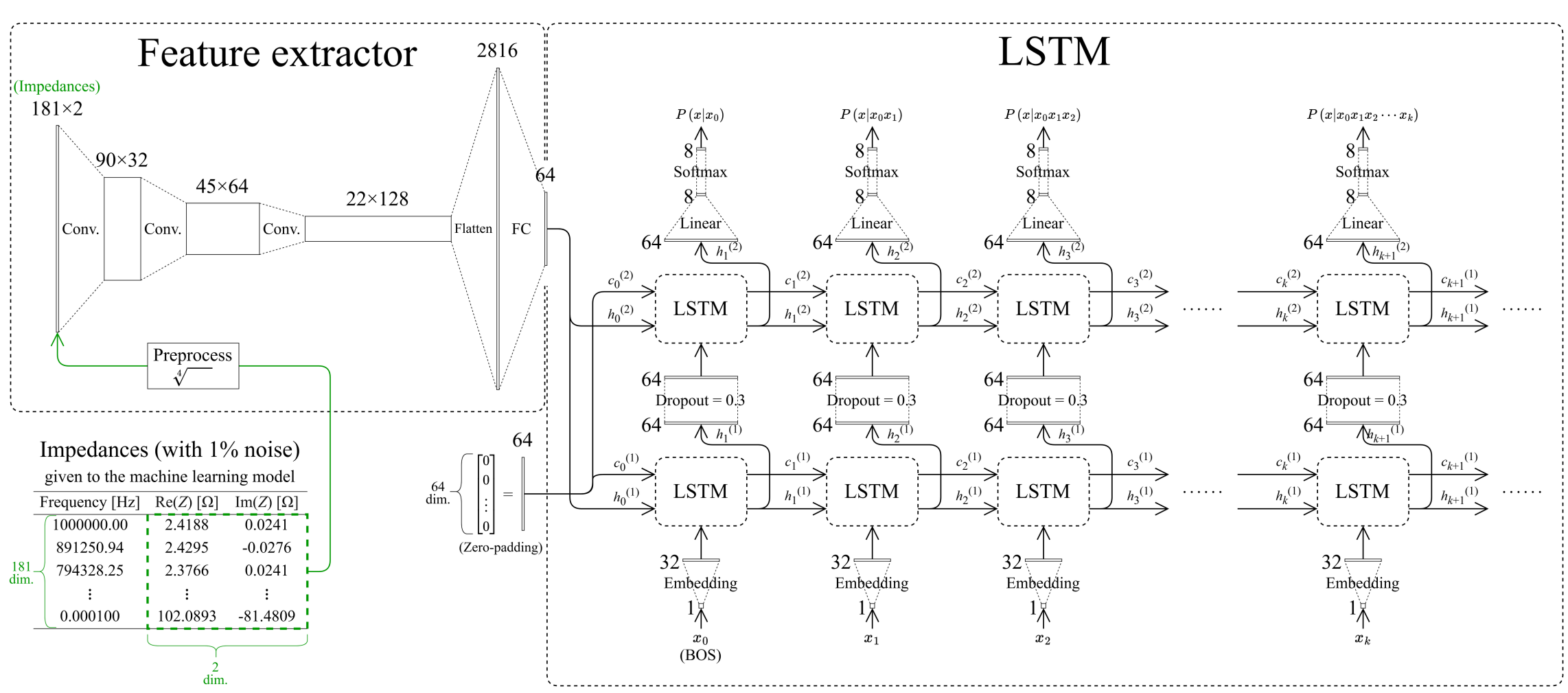

Fig. 2. Structure of the proposed neural network model using a feature extractor and an LSTM.

Table 1. Ids, symbols, and descriptions for ECM elements and connection.

| Id | Symbol | Description of symbol | Circuit diagram |
|---|---|---|---|
| 0 | (Blank) | For padding | - |
| 1 | (BOS) | BOS: Beginning of sentence | - |
| 2 | / | Parallel connection | - |
| 3 | R | Resistor | |
| 4 | C | Capacitor | |
| 5 | L | Inductor | |
| 6 | Q | Constant phase element | |
| 7 | W | Warburg impedance (semi-infinite diffusion) | |

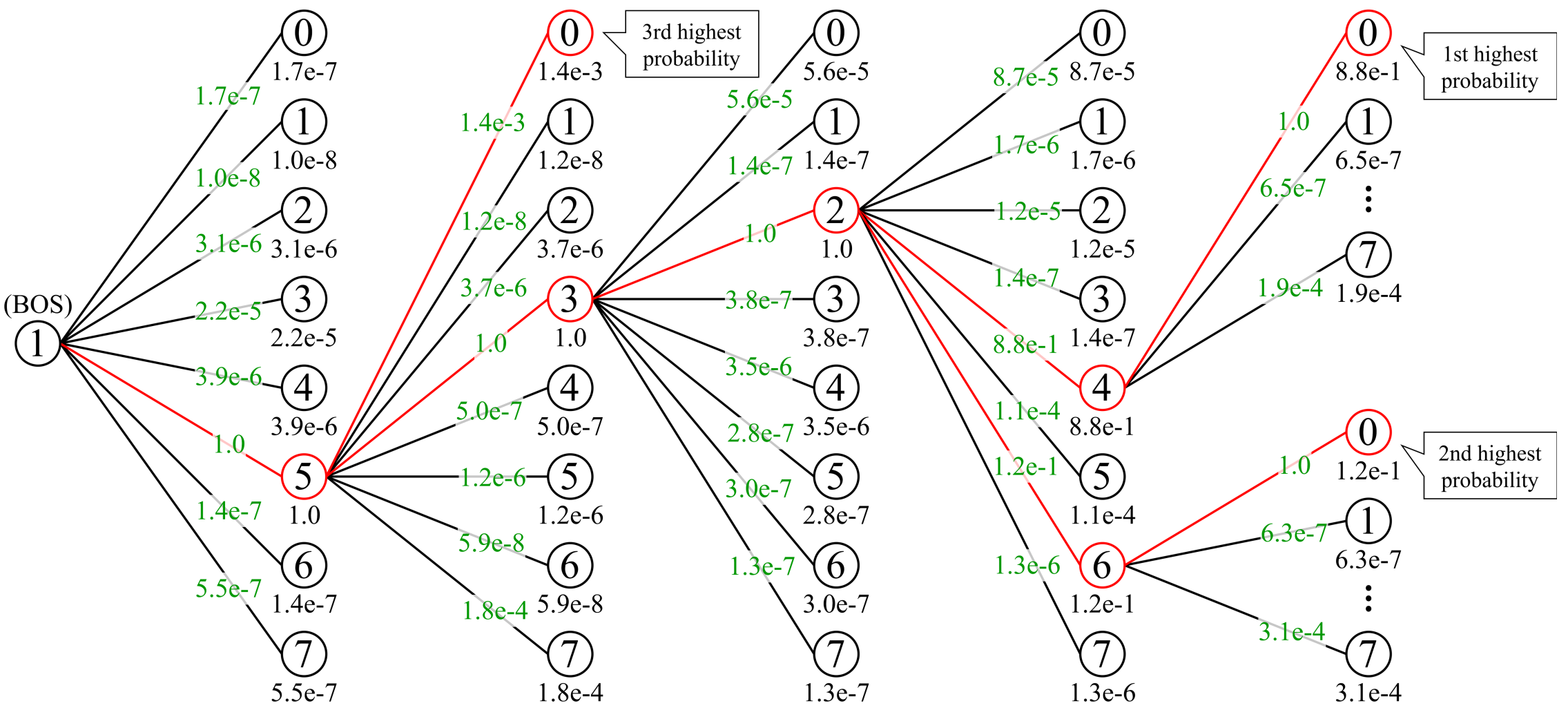

Fig. 3. Example of adaptive beam search to select the ECM candidates.

In this paper, we propose applying a combination of a feature extractor and LSTM as shown in Fig. 2, a structure that is often applied to image captioning [15][16]. This combination enables the ML model to automatically learn the relationship between an input impedance at swept frequency and an output string representing an appropriate ECM.

As a preprocessing step for ML, the real and imaginary parts of the impedance are fourth-root-transformed as in the following equation (1), where $Z$ is the original impedance, $\tilde{Z}$ is its fourth-root-transformed value, $\mathrm{Re}(Z)$ and $\mathrm{Im}(Z)$ are the real and imaginary parts of $Z$, respectively, $j$ is the imaginary unit, and $\alpha$ is a constant set to 0.25 in this paper.

$$\tilde{Z} = (\mathrm{Re}(Z))^{\alpha} + j \cdot (\mathrm{Im}(Z))^{\alpha} \tag{1}$$

This fourth-root-transforming preprocessing has the effect of mitigating the impacts of a large inductance and a large Warburg impedance and of making it easier for the ML model to analyze features in the mid-frequency range.

Each convolutional block in the feature extractor (denoted “Conv.” in Fig. 2) consists of five layers: a convolutional layer (with kernel size of 5, stride of 1, and padding of 2), a batch-normalization layer (with a momentum of 1.0 and an affine layer applied), a ReLU (Rectified Linear Unit) layer, a max-pooling layer (with a kernel size of 2 and a stride of 2), and a dropout later (with a dropout of 0.3). The combination of the three convolutional blocks works as a feature extractor, which extracts important features while reducing the amount of data [23]. The final feature map sized $22 \times 128$ is flattened and passed to the FC (Fully Connected) block. The FC block consists of five layers: a FC layer, a batch-normalization layer (with momentum of 1.0 and affine layer applied), a ReLU layer, a dropout later (with dropout of 0.3), and a second FC layer.

A stacked LSTM with two LSTM layers is applied after the feature extractor, and the output from the feature extractor consisting of 64 values is applied as the initial value of the hidden parameters for each LSTM layer. The initial cell parameters, sized 64, are padded with zeros, and the initial input to the stacked LSTM is the value designated as BOS (Beginning of Sentence). In the first step, these hidden, cell, and input values are passed to the stacked LSTM, and eight values as logits are obtained after passing through a linear stack after the stacked LSTM. These logits are normalized so that they range between 0 to 1 and sum to 1 by passing through a SoftMax layer. These normalized values satisfy the mathematical axioms of probability can be treated as probabilities, and are referred to simply as probabilities in the remainder of this paper; however, it must be noted that they merely represent confidence scores calculated by LSTM, not true statistical probabilities. After the probability for each word id is calculated, the word id with the highest probability is selected and then input to the stacked LSTM in the next step to obtain the next probabilities. Repeating this step provides the most probable combination of word ids.

The string representing the most possible ECM can be obtained from the word ids using Table 1.

For example, if the word ids “1532400…0” are obtained, the string representing the most probable ECM is “L-R/C,” inserting the symbol “-” representing series connection between element symbols unless the symbol “/” representing a parallel connection already existed, and tailing spaces represented by the word id 0 are ignored. It must also be noted that not only the word id with the highest probability but also another word id with second-, third-, or lower-ranked probability can be passed to the LSTM in the next step.

## 2.2 Adaptive beam search

There are cases where the correct ECM is not the most probable one but the second, third, or lower-ranked candidate. Therefore, in order to increase the likelihood of obtaining the true ECM, it is necessary to also retrieve the second-, third-, and lower-ranked ECMs from the stacked LSTM during the evaluation process. A standard LSTM typically employs a one-hot (greedy) strategy, in which only the character with the highest probability is selected at each generation step and lower possible ECM cannot be generated. As one of the solutions toward this issue, we propose applying relative threshold pruning strategy [22], a type of beam search, in the decoding process during the evaluation of a trained model.

Fig. 3 shows an example of adaptive beam search. Here, $p(n_k|n_0n_1n_2\ldots n_{k-1})$ in the $k$th step is the probability that the word id $n_k$ is appropriate following the sequence of word ids $n_0n_1n_2\ldots n_{k-1}$ and $p_{\mathrm{th}}$ is the threshold used in relative threshold pruning, set to 0.01 in this study. A string $n_0n_1n_2\ldots n_k$ is discarded when the following inequation (2) holds.

$$p(n_k|n_0n_1n_2\ldots n_{k-1}) < p_{\mathrm{th}} \cdot \max_n p(n|n_0n_1n_2\ldots n_{k-1}) \tag{2}$$

At the first step in Fig. 3, where BOS (word id 1) is input to the stacked LSTM, the model provides “5” as the most probable next word id, with the probability of $p(5|1) \approx 1.0$, while the other word ids provides other word ids with the probabilities lower than $p_{\mathrm{th}} \cdot p(5|1) = 0.010$ and discarded because inequation (2) holds. Therefore, only the sequence “15” remains under consideration and is input to LSTM at the second step. Similarly, at the second step, where “5” is input, provides “3” as the most probable next word id and “0” as the only word ids that do not satisfy inequation (2) and therefore are not discarded. Therefore, the word ids “153” and “150” remain under consideration, and in the third step two trials are conducted: one is to input “0” and another is to input “3”. However, we never discuss about inputting “0” since in this case the well-trained stacked LSTM in this research always provides “0” as the only word id that remains under consideration. The third step, where “3” is input, provides “2”, the fourth step provides “4” and “6”, and the fifth steps, where “4” and “6” are input, both provides only “0”. All strings under consideration in this example are “1500…0”, “1532400…0”, and “1532600…0.” The probability for each string is calculated with the following equation (3):

$$p(n_0n_1n_2\ldots n_{20}) = \prod_{k=1}^{20} p(n_k|n_0n_1\ldots n_{k-1}) \tag{3}$$

The probabilities are therefore calculated as $p(1500\ldots0) = 1.0 \cdot 1.4\text{e-}3 = 1.4\text{e-}3$ , $p(1532400\ldots0) = 1.0 \cdot 1.0 \cdot 1.0 \cdot 8.8\text{e-}1 \cdot 1.0 = 8.8\text{e-}1$ , and $p(1532600\ldots0) = 1.0 \cdot 1.0 \cdot 1.0 \cdot 1.2\text{e-}1 \cdot 1.0 = 1.2\text{e-}1$. Word ids translated into elements symbol using Table 1, the three most probable ECMs are “L-R/C”, “L-R/Q”, and “L”, in the descending order of probability. This adaptive beam search strategy is applied only in the testing process in this paper, not in the training process, in order to reduce computation time in the training process.

## 2.3 Generation of datasets

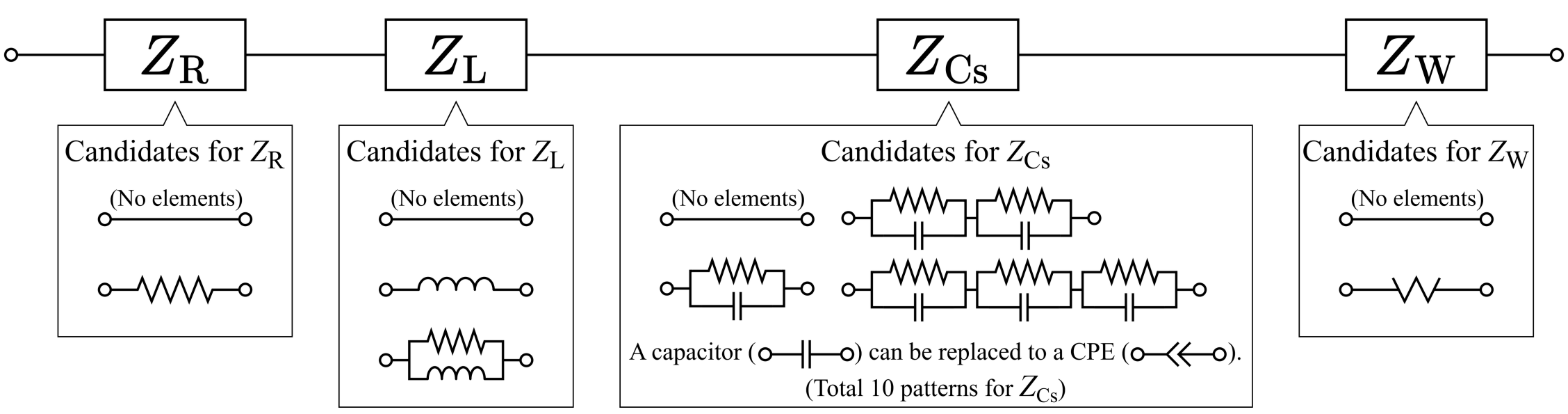

Fig. 4. Schematic diagram of the circuit construction for dataset generation

In addition to how to estimate an ECM, it is also necessary to consider how to generate the datasets used to train and evaluate the proposed ML model. Supervised learning requires large datasets each of which includes feature data and label data. In the training process, the ML model is given the feature data and outputs an estimated value of the label data, after which it is adjusted to output the true label data. In the evaluation process, how well the model outputs the true label data when the feature data is given is examined. We trained and evaluated the model with randomly generated circuits in order to demonstrate that the proposed methodology has the potential to estimate any ECM and is therefore applicable to any objects. In this paper, each dataset consists of the impedances of a randomly generated circuit as the feature data and a string representing the circuit as the label data.

To generate a dataset, first, a circuit topology was randomly selected from a total 119 candidates: $Z_{\mathrm{R}}$, $Z_{\mathrm{L}}$, $Z_{\mathrm{Cs}}$, and $Z_{\mathrm{W}}$ are selected uniformly at random from each list of candidates in Fig. 4 and connected in series each other. For example, if R, no elements, R/C, and no elements are selected for $Z_{\mathrm{R}}$, $Z_{\mathrm{L}}$, $Z_{\mathrm{Cs}}$, and $Z_{\mathrm{W}}$, respectively, the string representing a circuit "R-R/C" was generated, where "-" represents a series connection and "/" represents a parallel connection. This string is used as a label data in ML. If a circuit containing no elements was created, it was removed and an alternative circuit was generated in the same way. Second, the parameters of the elements in each selected circuit were log-uniformly or uniformly sampled according to the condition shown in Table 2. Then, with the frequency logarithmically swept from 0.1 mHz to 1.0 MHz over 101 points, the impedance of the circuit at each frequency was calculated using following equations (4) to (8):

$$Z_{\mathrm{R}} = R \tag{4}$$

$$Z_{\mathrm{L}} = j2\pi fL \tag{5}$$

$$Z_{\mathrm{C}} = \frac{1}{j2\pi fC} = -j\left(\frac{1}{2\pi fC}\right) \tag{6}$$

$$Z_{\mathrm{CPE}} = \frac{1}{(j2\pi f)^p \cdot T} = -j^p\left(\frac{1}{(2\pi f)^p \cdot T}\right) \tag{7}$$

$$Z_{\mathrm{W}} = \frac{(1-j)\sigma}{\sqrt{2\pi f}} \tag{8}$$

where $Z_{\mathrm{R}}$, $Z_{\mathrm{L}}$, $Z_{\mathrm{C}}$, $Z_{\mathrm{CPE}}$, and $Z_{\mathrm{W}}$ are the impedances of a resistor, an inductor, a capacitor, a CPE (constant phase element), and a Warburg impedance (semi-infinite diffusion), respectively, $R$, $L$, $C$, $T$, $p$, and $\sigma$ are the resistance, inductance, capacitance, magnitude of CPE, phase factor, and coefficient, respectively, and $f$ is the frequency. In addition, 1% white noise was added as shown in the following equation (9) to simulate measurement errors.

$$Z_{\mathrm{noisy}} = Z + \alpha|Z|(\eta_{\mathrm{Re}} + j\eta_{\mathrm{Im}}) \tag{9}$$

Here, $Z$ and $Z_{\mathrm{noisy}}$ are the impedances before and after adding the white noise, respectively, $|Z|$ is the magnitude of $Z$, $\alpha$ is the level of noise which is set to 0.01 in this study, and $\eta_{\mathrm{Re}}, \eta_{\mathrm{Im}} \sim N(0,1)$ are independent random variables sampled from a standard normal distribution. This noisy impedance was used as the feature data in ML.

# 3. Evaluation of the proposed methodology

Table 2. Value range of the circuit elements for impedance calculation to generate dataset

| Parameter | Range | Distribution |
|---|---|---|
| Resistance $R$ | $10^0$ – $10^1$ [Ω] | Log-uniform |
| Inductance $L$ | $10^{-2}$ – $10^{-4}$ [H] | Log-uniform |
| Capacitance $C$ | $10^{-2}$ – $10^{-4}$ [F] | Log-uniform |
| Magnitude $T$ | $10^{-2}$ – $10^{-4}$ $[\mathrm{F} \cdot \mathrm{s}^{p-1}]$ | Log-uniform |
| Phase factor $p$ | 0.6 – 0.9 | Uniform |
| Coefficient $\sigma$ | $10^0$ – $10^2$ $[\Omega \cdot \mathrm{s}^{0.5}]$ | Log-uniform |

Table 3. Conditions to train the machine learning model.

| Parameter | Value |
|---|---|
| Training epochs | 50 epochs |
| Batch size | 128 |
| Loss function | Cross entropy loss |
| Optimizer | Adam |
| Learning rate | 0.001 |
| Maximum gradient norm | 1.0 |
| Number of datasets | 200,000 (100,000 for training, 100,000 for testing) |
| Threshold for relative threshold pruning | $p_{\mathrm{th}} = 0.01$ |

This section describes the experimental conditions and evaluation metrics used to assess the proposed methodology. The evaluation has two objectives: to demonstrate that the proposed methodology can estimate a wide variety of ECMs, and to verify the effect of the fourth-root-transforming preprocess.

The proposed methodology was trained and evaluated on the generated datasets under the conditions shown in Table 3 and described in section 2.1. The evaluation was conducted as follows. First, the ML model was trained without adaptive beam search; that is, it was trained to return only a single ECM to match the target label. The adaptive beam search strategy was not applied in the training process so as not to increase the computation time required for the training, but was applied in the testing process in order to examine not only the first but also the second and lower-ranked probable ECMs. For the evaluation of each dataset after the training was completed, at most five ECMs with the highest probabilities were extracted; the upper limit of five was chosen because, in preliminary experiments, the correct ECM rarely appeared beyond the fifth candidate, so extracting the further candidates provided little additional benefit. The evaluation result was then labelled according to which ECM the target label matched: it was labeled "1st" when it matched the most probable ECM, "2nd" when it matched the second most probable ECM, "$n$th" when it matched the $n$th most probable ECM, and "Not matched" when it did not match any of the (at most five) ECMs.

To verify the effect of the fourth-root-transforming preprocess, the same datasets were evaluated both with and without the preprocess. In the evaluation without the preprocess, the fourth-root-transforming preprocess in Fig. 2 was applied in neither the training nor the testing process of ML, whereas in the evaluation with the preprocess, it was applied in both. Comparing the two reveals the effect of the fourth-root-transforming preprocess, while the results obtained with the preprocess also serve to evaluate the performance of the proposed methodology itself.

In addition, to confirm the low computational cost claimed in section 1, we measured the time required to run the evaluation under a batch size of 1 and calculated the average and maximum time required per dataset. This measurement was performed only for the evaluation with the preprocessing, since the fourth-root-transforming preprocess itself incurs a negligible computational cost and thus the measured time is representative of both cases.

# 4. Results and discussion

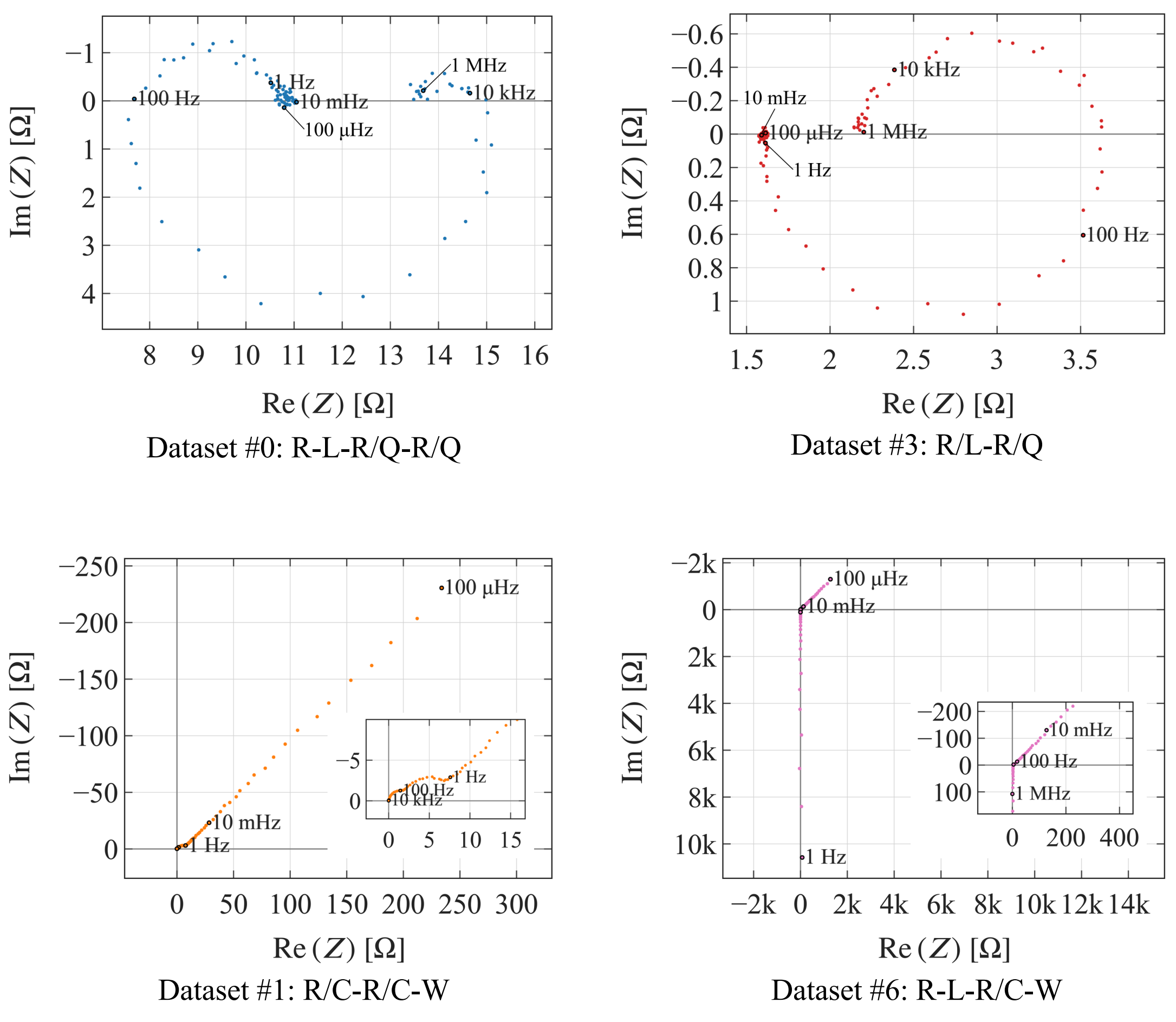

Fig. 5. Examples of the generated impedances with raw values

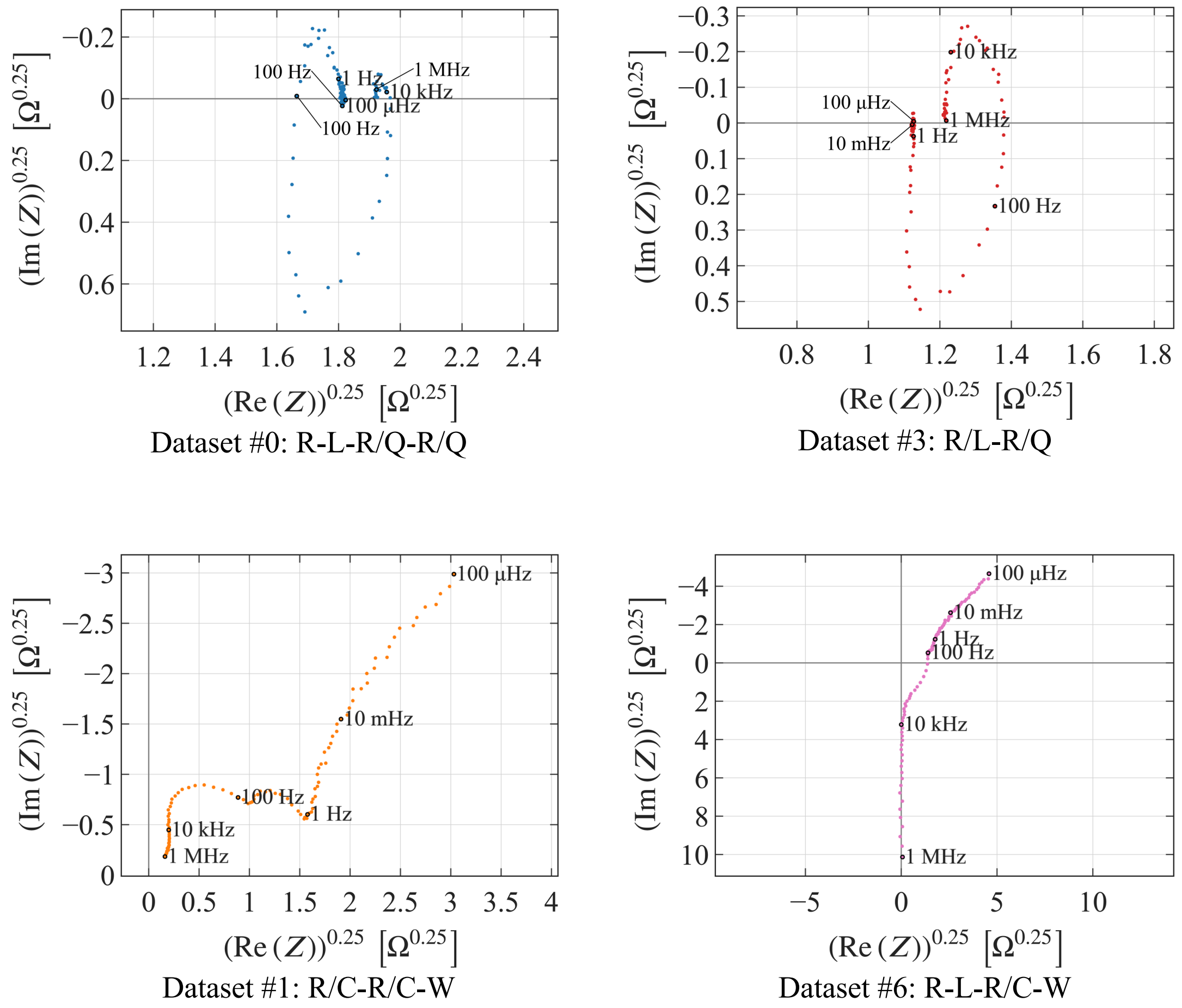

Fig. 6. Examples of the preprocessed impedances

Table 4. Summary on the evaluation results

| Result label | Count (without preprocess) | Count (with preprocess) |
|---|---|---|
| 1st | 39444 | 77823 |
| 2nd | 16943 | 13609 |
| 3rd | 12144 | 4336 |
| 4th | 8296 | 2037 |
| 5th | 6022 | 1001 |
| Not matched | 17151 | 1194 |
| Total | 100000 | 100000 |

Table 5. Actual ECMs and estimation results without the preprocess

| Label | Dataset #0 | Dataset #1 | Dataset #15 | Dataset #885 |
|---|---|---|---|---|
| Actual ECM and Nyquist plot | R-R/L-R/Q-R/Q | R/C-R/C-W | R/C-R/C-R/Q | L-R/C-R/C |
| 1st most possible ECM | R-R/L-R/C (24.0%) | R/C-R/C-W (36.6%) | R/Q-R/Q-R/Q (35.6%) | R-L-R/C (28.0%) |
| 2nd most possible ECM | R-R/L-R/Q (20.3%) | R/C-W (13.6%) | R/C-R/Q-R/Q (26.2%) | R-L-R/Q (14.0%) |
| 3rd most possible ECM | R-R/L-R/C-R/Q (16.7%) | R/C-R/Q-W (11.5%) | R/C-R/C-R/Q (15.0%) | L-R/C-R/Q (13.8%) |
| 4th most possible ECM | R-R/L-R/Q-R/Q (9.7%) | R/Q-W (10.4%) | R/Q-R/Q (12.8%) | L-R/Q (10.8%) |
| 5th most possible ECM | R-R/L-R/C-R/C (5.6%) | R/C-R/C-R/C-W (9.6%) | R/C-R/Q (4.676%) | L-R/Q-R/Q (8.7%) |

Note: the values in parentheses are the probabilities (confidential scores) by the ML model.

Table 6. Actual ECMs and estimation results with the preprocess

| Label | Dataset #0 | Dataset #1 | Dataset #15 | Dataset #885 |
|---|---|---|---|---|
| Actual ECM and Nyquist plot | R-R/L-R/Q-R/Q | R/C-R/C-W | R/C-R/C-R/Q | L-R/C-R/C |
| 1st most possible ECM | R-R/L-R/Q-R/Q (94.1%) | R/C-R/C-W (91.1%) | R/C-R/Q-R/Q (51.5%) | L-R/Q (27.8%) |
| 2nd most possible ECM | R-R/L-R/Q-R/Q-R/Q (4.7%) | R/C-R/Q-W (8.7%) | R/C-R/Q (21.6%) | L-R/Q-R/Q (27.1%) |
| 3rd most possible ECM | - | - | R/C-R/C-R/Q (18.9%) | R-L-R/C (18.4%) |
| 4th most possible ECM | - | - | R/Q-R/Q-R/Q (4.8%) | L-R/C-R/Q (14.4%) |
| 5th most possible ECM | - | - | R/Q-R/Q (3.1%) | R-L-R/Q (10.6%) |

Note: the values in parentheses are the probabilities (confidential scores) by the ML model.

Fig. 5 shows four typical examples of impedance spectra from the generated datasets. The datasets include not only practical spectra for which electrochemists can easily estimate an ECM but also some spectra for which estimating an ECM is difficult. Fig. 5 indicates that an important feature appearing in mid-frequency range is inconspicuous, especially in spectra with large impedances caused by a single inductance and a Warburg impedance such as those of dataset #1 and #6. Fig. 6 shows the fourth-root-transformed impedances of the four impedance spectra, in which the mid-frequency features that are difficult to see in Fig. 5 become clear. This indeicate that the fourth-root-transforming preprocess has the effect of mitigating the impacts of a large inductance and a large Warburg impedance, which enables the ML model to learn the mid-frequency features.

Table 4 shows the summary of the evaluation results without and with the fourth-root-transforming preprocess using the same 100,000 datasets for testing. In the evaluation with the preprocess, the true label matched the most possible ECM in 77.8% of the datasets. The adaptive beam search strategy provides not only the first but also the second, third, or lower-ranked probable ECMs, and the true label matched one of the five most probable ECMs in 98.8% of the datasets. Without the fourth-root-transforming preprocess, these proportions declined to 4.8% and 13.3%, respectively, indicating that the preprocess helps the ML model capture the feature of impedances in the mid-frequency range.

Another notable result concerns the computation time on the evaluation with the fourth-root-transforming preprocess. The evaluation of each dataset took no longer than 17.8 milliseconds on average, with a batch size of 1, on a laptop equipped with an 8-core Intel Ultra 7 processor with 32 GB of RAM running Windows 11, whereas previous research [14] reports that GEP usually requires 2 to 10 hours for an estimation on a single dataset, using a laptop equipped with an 8-core Intel i7 processors and 16 GB of RAM. This estimation time is far shorter than that reported in the previous research and short enough that the users can wait for estimation to complete without stress.

Table 5 shows four typical examples of the evaluation results with the fourth-root-transforming preprocess. For many of the datasets, the estimation was successful even when the impedance was strongly affected by a large inductance or a large Warburg impedance, as in Dataset #1. Moreover, these results show much higher ranks and much higher probabilities compared with the results without the preprocess as shown in Table 6, which indicates that the fourth-root-transforming preprocess substantially improved the accuracy of the ML model by mitigating the large impacts of large inductances and large Warburg impedances. In some cases, such as Dataset #15, the true ECM was generated not as the most probable but the second- or lower-ranked probable ECM. Even in such cases, the true ECM can be found by applying parameter fitting to each ECM generated by the ML model.

Three problems remain to be solved to achieve the goal of this research: automatically estimate an appropriate ECM for any objects. First, ECMs similar to the true one are sometimes generated instead of the true ECM itself. For Dataset #885 in the evaluation with the preprocess, for example, the true ECM "L-R/C-R/C" could not be generated, whereas ECMs similar to it, such as "L-R/Q-R/Q" or "L-R/C-R/Q", were generated. Second, the methodology must be adapted to other circuits not yet considered in this paper, including not only circuits with more resistive-capacitive or resistive-inductive elements but also circuits that cannot be simplified into series or parallel topologies, such as bridge circuits and non-planar circuits. Third, it is also necessary to estimate multiple ECMs since it has been reported that a given circuit can have other circuits with identical impedances [2].

# 5. Conclusion

In this research, we proposed a novel methodology for an ECM estimation that combines a feature extractor and an LSTM network. The central idea is to represent an ECM as a serialized string of symbols denoting circuit elements and their series and parallel connections, and to have the LSTM generate this string directly from the measured impedance. This approach is fundamentally different from existing methods. Because the LSTM inherently handles variable-length sequential data, the proposed method generates an ECM directly, without any parameter fitting during estimation and without any prior assumption about the number of circuit elements. To our knowledge, this is the first demonstration that an ECM can be estimated by directly generating topology itself with a NN.

To improve the estimation accuracy, we introduced a fourth-root transformation as a preprocessing step, which mitigates the dominant impacts such as large inductances and Warburg impedances and thereby reveals the features in the mid-frequency range that are essential for distinguishing ECMs. We also adopted an adaptive beam search strategy in the decoding process, which yields not only the single most probable ECM but also lower-ranked candidates.

The evaluation process using the 100,000 datasets proved that the proposed methodology successfully answered a correct circuit topology as the most probable ECM for the 77.8% of the datasets and answered as one of the most five probable ECMs for the 98.8% of the datasets within the average calculation time of 17.8 milliseconds for the evaluation of each dataset, even though 1% white noise is added to impedance before giving it to the proposed ML model. The remaining issues include improving accuracy, adapting also to other ECM topologies not yet considered in this paper, and estimating multiple ECMs which have identical impedances.